\documentclass[preprint]{elsarticle}
\usepackage{graphicx}
\usepackage{amsmath}
\usepackage{amssymb}
\usepackage[nolists]{endfloat}

\newcommand{\half}{\frac{1}{2}}

\newcommand{\partiald}[2]{\frac{\partial #1}{\partial #2}}
\newcommand{\totald}[2]{\frac{d #1}{d #2}}
\newcommand{\totaldd}[2]{\frac{d^2 #1}{d #2^2}}

\newcommand{\bq}{\mathbf{q}}

\journal{Journal of Computational Physics}

\begin{document}

\begin{frontmatter}

\title{Comment on ``Symplectic integration of magnetic systems": a proof that the Boris algorithm is not variational}
\author[pppl]{C. L. Ellison \fnref{c}}
\ead{lellison@pppl.gov}
\author[pppl]{J. W. Burby}
\author[pppl,ustc]{H. Qin}

\address[pppl]{Princeton Plasma Physics Laboratory, Princeton, NJ 08543, USA}
\address[ustc]{Department of Modern Physics, University of Science and Technology of China, Hefei, Anhui 230026, China}

\fntext[c]{\copyright 2015. This manuscript version is made available under the CC-BY-NC-ND 4.0 license http://creativecommons.org/licenses/by-nc-nd/4.0/}


\end{frontmatter}

\section{Introduction}
\label{sec:introduction}

The Boris algorithm is a popular technique for the numerical time advance of charged particles interacting with electric and magnetic fields according to the Lorentz force law \cite{Boris_1970, Birdsall_1985_Boris, Stoltz_2002, Penn_2003}. Its popularity stems from simple implementation, rapid iteration, and excellent long-term numerical fidelity \cite{Boris_1970, Qin_2013}. Excellent long-term behavior strongly suggests the numerical dynamics exhibit conservation laws analogous to those governing the continuous Lorentz force system \cite{Hairer_2006}. Without conserved quantities to constrain the numerical dynamics, algorithms typically dissipate or accumulate important observables such as energy and momentum over long periods of simulated time \cite{Hairer_2006}. Identification of the conservative properties of an algorithm is important for establishing rigorous expectations on the long-term behavior; energy-preserving, symplectic, and volume-preserving methods each have particular implications for the qualitative numerical behavior \cite{Hairer_2006, Xia_1992, Zaijiu_1994, Quispel_1995, Quispel_1998, He_2015}. 

A recent Journal of Computational Physics article \cite{Webb_2014} argued that the Boris algorithm is a variational integrator \cite{Marsden_2001} and effectively equivalent to a symplectic integration of Hamilton's equations for the charged particle dynamics. However, this claim contradicts claims made in Refs.\,\cite{Qin_2013, Zhang_2014}, which demonstrate the volume-preserving property of the Boris scheme. Although symplectic integrators are also volume preserving, volume-preserving integrators are not necessarily symplectic, and Ref.\,\cite{Qin_2013} concludes the Boris algorithm is \emph{only} volume-preserving based on direct calculation of a symplecticity condition. Interestingly, the calculations in Ref.\,\cite{Qin_2013} do \emph{not} preclude a discrete variational formulation, as claimed in Ref.\,\cite{Webb_2014}. The reason for this is that there exists freedom in the identification of a discrete momentum $\mathbf{p}_k$ corresponding to a particular state with position $\mathbf{q}_k$ and velocity $\mathbf{v}_k$. The calculations in Ref.\,\cite{Qin_2013} show that the Boris scheme is not symplectic when $\mathbf{p}_k = m \mathbf{v}_k$ or $\mathbf{p}_k = m \mathbf{v}_k + \frac{e}{c} \mathbf{A}(\mathbf{q}_k)$, where $m$ is the particle's mass, $e$ its charge, $c$ the speed of light and $\mathbf{A}$ the magnetic vector potential. For variational integrators, the discrete momenta depend on the particular choice of discrete Lagrangian, and would likely differ from the preceding two definitions. There therefore exists a controversy in the literature, and identifying the conservation properties of the Boris method requires resolving this discrepancy.

In this letter, we prove that the Boris algorithm does not possess a discrete variational formulation by demonstrating violation of a discrete Helmholtz condition \cite{Bourdin_2013}. The discrete Helmholtz condition is a necessary and sufficient requirement for second-order finite-difference equations to be variational algorithms. The argument for the variational formulation of Boris presented in Ref.\,\cite{Webb_2014} therefore cannot be correct, and we detail two invalidating flaws in the argument. The first flaw involves improper calculation of the discrete Euler-Lagrange equations corresponding to the specified discrete action. The second flaw is the introduction of truncation error after performing the discrete action minimization, as identified as suspect in Ref.\,\cite{Zhang_2014}. Ultimately, the Boris algorithm has only been demonstrated to be volume-preserving, and the long-term behavior should be attributed to this property.

\section{The Boris algorithm is not a variational integrator}
\label{sec:the_boris_algorithm_is_not_a_variational_integrator}

A popular technique for constructing conservative numerical algorithms is to discretize the variational principle underlying the equations of motion \cite{Marsden_2001}. By performing all approximations at the level of the action principle, the numerical update inherits several conservation laws of the continuous dynamics, including a discrete Noether's theorem. Centrally, provided certain regularity and consistency conditions are satisfied, performing a time advance using a variational integrator is equivalent to using a symplectic integrator on Hamilton's equations \cite{Marsden_2001}, and is therefore known to exhibit bounded energy error for exponentially long times \cite{Hairer_2006}. 

Given a finite-difference approximation of an Euler-Lagrange equation, then, it is of interest to determine whether the algorithm constitutes a variational integrator. In the continuous setting, determining whether a second-order differential equation may be derived as the Euler-Lagrange equations corresponding to an action principle is known as ``Helmholtz's inverse problem of the calculus of variations" \cite{Helmholtz_1887, Douglas_1941}. Recent work has established a discrete analog to the solution of this problem for determining whether a given finite-difference method for second-order differential equations may be derived from a discrete action principle, i.e. is a variational integrator \cite{Bourdin_2013}. This section applies the discrete Helmholtz condition to the Boris algorithm, and finds that it is only satisfied when the magnetic field $\mathbf{B}(\mathbf{q}, t)$ is a constant. 

A second-order differential equation is specified by a function $\mathbf{f}(\mathbf{q}, \mathbf{v}, \mathbf{a}, t)$ and the condition that 
\begin{equation}
\mathbf{f}(\bq(t), \totald{\bq}{t}(t), \totaldd{\bq}{t}(t), t) = 0, \quad \forall\, t = [0, T] \subset \mathbb{R}.
\end{equation}
Suppose we want to approximate the solution to the differential equation $\bq(t)$ at discrete 
instances in time $t_k = 0, h, ..., N h$, with $N h = T$. A second-order finite-difference equation is specified by a function $\mathbf{P}(\bq, \mathbf{v}_{-}, \mathbf{v}_{+}, \mathbf{a}, t, \xi)$ and the condition that:
\begin{equation}	
	\mathbf{P}(\bq_k, \frac{\bq_{k} - \bq_{k-1}}{h}, \frac{\bq_{k+1} - \bq_k}{h}, \frac{\bq_{k+1} - 2 \bq_k + \bq_{k-1}}{h^2}, t_k, h) = 0, \quad \forall\, k = 1, ..., N-1,
\end{equation}
where $\bq_k$ is the numerical approximation to the solution $\bq(t=t_k)$. Following Ref.\,\cite{Bourdin_2013}, the finite-difference equation is a variational integrator if and only if $\mathbf{P}$ satisfies:
\begin{equation}
	\frac{1}{h}\left(\partiald{\mathbf{P}}{\mathbf{a}}(\star_k) - \partiald{\mathbf{P}}{\mathbf{a}}(\star_{k-1}) \right) = \partiald{\mathbf{P}}{\mathbf{v}_{+}}(\star_k) + \partiald{\mathbf{P}}{\mathbf{v}_{-}}(\star_{k-1}), \quad \forall\, k = 2, ..., N-1, 
	\label{eq:discrete_helmholtz_condition}
\end{equation}
where $\star_k = (\bq_k, \frac{\bq_{k} - \bq_{k-1}}{h}, \frac{\bq_{k+1} - \bq_k}{h}, \frac{\bq_{k+1} - 2 \bq_k + \bq_{k-1}}{h^2}, t_k, h)$, and the condition must hold for arbitrary $(\bq_{k}, t_k)$ for all $k$. Equation (\ref{eq:discrete_helmholtz_condition}) is known as the \emph{discrete Helmholtz condition}. Note that although the condition in Ref.\,\cite{Bourdin_2013} presumes a one dimensional system for notational clarity, we verified the necessity of this condition in higher dimensions. 

The Boris algorithm may be written as \cite{Boris_1970}:
\begin{equation}
	\frac{\bq_{k+1} - 2 \bq_k + \bq_{k-1}}{h^2} = \left( \frac{\bq_{k+1} - \bq_{k-1} }{2 h}\right) \times \mathbf{B}(\bq_k) + \mathbf{E}(\bq_k).
\end{equation}
So, the Boris algorithm is a second-order finite-difference method where the $i$-th component of the vector-valued function $\mathbf{P}$ is given by:
\begin{equation}
	P^i(\bq, \mathbf{v}_{-}, \mathbf{v}_{+}, \mathbf{a}, t, \xi) = a^i - \epsilon^{i}_{jl} \left(\frac{v^j_{+} - v^j_{-}}{2} \right) B^l(\mathbf{q}, t) - E^i(\mathbf{q}, t), 
\end{equation}
where $\epsilon$ is the Levi-Civita symbol, and summation over repeated indices is implied. Evaluation of the terms in the discrete Helmholtz condition Eq.\,(\ref{eq:discrete_helmholtz_condition}) yields the condition that:
\begin{equation}
	0 = -\half \epsilon^i_{jl} \left(B^l(\bq_k, t_k) - B^l(\bq_{k-1}, t_{k-1}) \right),
\end{equation}
for all $i$ and $j$ equal to $1,2$ or $3$, and for arbitrary $(\bq_{k-1}, \bq_{k}, t_{k-1}, t_k)$ in $\mathbb{R}^{8}$. We can see, then, that the Boris algorithm admits a variational formulation only in the case when $\mathbf{B}(\bq,t)$ is a constant function of position and time. 

\section{Critique of the argument in Ref.\,\cite{Webb_2014}}
\label{sec:critique_of_the_argument}

Because the Boris algorithm violates the discrete Helmholtz condition, the claim in Ref.\,\cite{Webb_2014} that the scheme is a variational integrator cannot be correct except in the case of constant magnetic field $\mathbf{B}(\bq,t)$. The first misstep in the argument of Ref.\,\cite{Webb_2014} involves mis-calculation of the discrete Euler-Lagrange equations corresponding to the specified discrete action. In short, only the first component of a two-step map is presented. Although this is a critical flaw for the particular argument as presented, we find that the discrete Euler-Lagrange equations as presented in Ref.\,\cite{Webb_2014} are indeed variational, and we identify a different discrete action yielding these discrete Euler-Lagrange equations. The second critique is the introduction of truncation error after derivation of the discrete Euler-Lagrange equations, and we investigate the claim that the approximation may be considered equivalent to re-definition of the electric and magnetic field quantities. We find that this claim only holds in the case when the magnetic field $\mathbf{B}(\bq,t)$ is a constant, consistent with the result of Section \ref{sec:the_boris_algorithm_is_not_a_variational_integrator}. 

\subsection{Incomplete calculation of the discrete Euler-Lagrange equations}
\label{ssec:incomplete_calculation_of_the_discrete_euler-lagrange_equations}


In constructing a method for non-relativistic charged particle dynamics and ultimately arguing for the symplecticity of the Boris update, Ref.\,\cite{Webb_2014} approximates a small time interval of the continuous action by specifying a pair of discrete Lagrangians. Given a charged particle of mass $m$ and charge $e$ moving in electromagnetic fields described by normalized vector potential $\mathbf{a} = \frac{e}{mc}\mathbf{A}$ and scalar potential $\varphi = \frac{e}{m}\phi$, Ref.\,\cite{Webb_2014} specifies the following discrete Lagrangians:
\begin{align}
  L_d^1(\mathbf{q}_k^0, \mathbf{q}_k^1, t_k, h) & = \frac{1}{2} \frac{\|\mathbf{q}_k^1 - \mathbf{q}_k^0\|^2}{h/2} +  \frac{1}{2} \left(\mathbf{a}(\mathbf{q}_k^0, t_k) + \mathbf{a}(\mathbf{q}_k^1, t_k + h/2) \right) \cdot \left(\mathbf{q}_k^1 - \mathbf{q}_k^0 \right) - \nonumber \\
& \quad \frac{h}{2} \varphi(\mathbf{q}_k^1, t_k + h/2)    \label{eq:discrete_lagrangian_one} \\
  L_d^2(\mathbf{q}_k^1, \mathbf{q}_k^2, t_k, h) & = \frac{1}{2} \frac{\|\mathbf{q}_k^2 - \mathbf{q}_k^1\|^2}{h/2} + \frac{1}{2} \left(\mathbf{a}(\mathbf{q}_k^1, t_k + h/2) + \mathbf{a}(\mathbf{q}_k^2, t_k + h) \right) \cdot \left(\mathbf{q}_k^2 - \mathbf{q}_k^1 \right) -  \nonumber \\
& \quad \frac{h}{2} \varphi(\mathbf{q}_k^1, t_k + h/2).   \label{eq:discrete_lagrangian_two}
\end{align}
The corresponding discrete action is given by:
\begin{equation}
  S_d(\mathbf{q}^0_0, \mathbf{q}^1_0, \mathbf{q}^2_0=\mathbf{q}^0_1, \ldots, \mathbf{q}^2_N) = \sum_{k=0}^{N-1} L_d^1(\mathbf{q}_k^0, \mathbf{q}_k^1, t_k) + L_d^2(\mathbf{q}_k^1, \mathbf{q}_k^2, t_k),
\label{eq:discrete_action} 
\end{equation}
where the dependence on numerical step size $h$ is now implied and the identification $\mathbf{q}_k^2 = \mathbf{q}_{k+1}^0$ is made to link between steps. The discrete action is extremized by trajectories that satisfy the following discrete Euler-Lagrange equations:
\begin{align}
  D_2 L_d^1(\mathbf{q}_k^0, \mathbf{q}_k^1, t_k) + D_1 L_d^2(\mathbf{q}_k^1, \mathbf{q}_k^2, t_k) & = 0 \label{eq:general_del_one}\\
	D_2 L_d^2(\mathbf{q}_k^1, \mathbf{q}_k^2, t_k) + D_1 L_d^1(\mathbf{q}_{k+1}^0, \mathbf{q}_{k+1}^1, t_{k+1}) & = 0.
	\label{eq:general_del_two}
\end{align}
Here, $D_i$ is the slot derivative indicating differentiation with respect to the $i$-th argument. These equations correspond to Eqs.\,(13a, 13b) in Ref.\,\cite{Webb_2014}. For the specific discrete Lagrangians defined in Eqs.\,(\ref{eq:discrete_lagrangian_one}-\ref{eq:discrete_lagrangian_two}), the discrete Euler-Lagrange equations become:
\begin{align} 
 \left( \frac{\mathbf{q}_k^1 - \mathbf{q}_k^0}{h/2} - \frac{\mathbf{q}_k^2 - \mathbf{q}_k^1}{h/2}\right) + \frac{1}{2}\left(\mathbf{a}(\mathbf{q}_k^0, t_k) - \mathbf{a}(\mathbf{q}_k^2, t_k +h) \right) & + \nonumber  \\ 
 \frac{1}{2} \, \nabla \mathbf{a}(\mathbf{q}_k^1, t_k + h/2) \cdot \left( \mathbf{q}_k^2 - \mathbf{q}_k^0 \right) - h \, \nabla \varphi(\mathbf{q}_k^1, t_k  + h/2) & = 0  \label{eq:explicit_del_one}\\
 \left(\frac{\mathbf{q}_{k}^2 - \mathbf{q}_k^1}{h/2} - \frac{\mathbf{q}_{k+1}^1 - \mathbf{q}_k^2}{h/2} \right) + \frac{1}{2}\left( \mathbf{a}(\mathbf{q}_k^1, t_k + h/2) - \mathbf{a}(\mathbf{q}_{k+1}^1, t_{k+1} + h/2) \right) & + \nonumber \\
 \frac{1}{2} \, \nabla \mathbf{a}(\mathbf{q}_k^2, t_k + h) \cdot \left( \mathbf{q}_{k+1}^1 - \mathbf{q}_k^1 \right) & = 0 . \label{eq:explicit_del_two}
\end{align}

Equation (23) in Ref.\,\cite{Webb_2014} is presented as the set of discrete Euler-Lagrange equations corresponding to the discrete action in Eq.\,(\ref{eq:discrete_action}). However, comparison with Eqs.\,(\ref{eq:explicit_del_one}-\ref{eq:explicit_del_two}) reveals that this is incorrect. Equation (23) in Ref.\,\cite{Webb_2014} represents only the first component of a two-step map, namely Eq.\,(\ref{eq:explicit_del_one}) or Eq.\,(\ref{eq:general_del_one}). Trajectories generated by iterating Eq.\,(\ref{eq:explicit_del_one}) alone do not extremize the discrete action given in Eq.\,(\ref{eq:discrete_action}), so a discrete variational formulation of the Boris update cannot be concluded from the proposed discrete action.


While the preceding criticism is sufficient to invalidate the specific argument presented in Ref.\,\cite{Webb_2014} for the discrete variational formulation of the Boris scheme, an important question is whether inclusion of the full set of discrete Euler-Lagrange equations or selection of a slightly different discrete action could repair the argument. Indeed, if one checks the discrete Helmholtz condition on Eq.\,(\ref{eq:explicit_del_one}), one finds that it is \emph{satisfied}. A variational formulation for Eq.\,(\ref{eq:explicit_del_one}) exists, specifically with a discrete Lagrangian and action given by:
\begin{align}
	L_d(\bq_k, \bq_{k+1}, t_k) & = \frac{\|\bq_{k+1} -\bq_k\|^2}{2 h} + \half \left(\mathbf{a}(\mathbf{q}_k) + \mathbf{a}(\bq_{k+1}) \right) \cdot \left(\bq_{k+1} - \bq_k\right) - \nonumber \\
	 & \quad h\,\varphi(\bq_k, t_k) \\
	S_d(\bq_0, ..., \bq_N) & = \sum_{k=0}^{N-1} L_d(\bq_k, \bq_{k+1}, t_k).
\end{align}
To resolve the conflict with the result of Section $\ref{sec:the_boris_algorithm_is_not_a_variational_integrator}$, we turn to the second critique.

\subsection{Truncation of the discrete Euler-Lagrange equations}
\label{ssec:truncation_of_the_discrete_euler-lagrange_equations} 

Proofs of the structure-preserving properties of variational integrators utilize the discrete variational principle and therefore do not necessarily apply to approximations of the discrete Euler-Lagrange equations. Any introduction of truncation error after performing the action extremization procedure is liable to violate the formal conservation properties, so the variational nature of the truncated equations must be independently established. In the case of the Boris algorithm, the calculation in Section \ref{sec:the_boris_algorithm_is_not_a_variational_integrator} demonstrates this is not possible.

In the process of relating the familiar Boris scheme to variational integration of charged particle dynamics, Ref.\,\cite{Webb_2014} introduces the following approximation:
\begin{equation}
  \frac{1}{2}\left(\mathbf{a}(\mathbf{q}^0_k, t_k) -  \mathbf{a}(\mathbf{q}^2_k, t_{k} + h) \right) \approx -\frac{1}{2} (\mathbf{q}^2_{k} - \mathbf{q}_k^0) \cdot \nabla \mathbf{a}(\mathbf{q}_k^1, t_k + h/2) - \frac{h}{2} \frac{\partial \mathbf{a}}{\partial t}(\mathbf{q}_k^1, t_k + h/2). 
  \label{eq:vector_potential_approximation}
\end{equation}
While Ref.\,\cite{Webb_2014} correctly argues that such a truncation does not affect the local order of accuracy of the method, the justification that the approximated equations retain a variational formulation is insufficient. The Reference identifies that justification is required, and states that the approximation may be instead interpreted as a re-definition of the electric and magnetic fields. Presumably, this means that functions $\widetilde{\mathbf{B}}(\bq,t), \widetilde{\mathbf{E}}(\bq,t)$ may be defined such that:
\begin{align}
	\frac{(\bq_{k+1} - \bq_{k-1})}{2 h} \times \widetilde{\mathbf{B}}(\bq_k, t_k) + \widetilde{\mathbf{E}}(\bq_k, t_k) & = \frac{1}{2h}\left(\mathbf{a}(\bq_{k-1}, t_{k-1}) - \mathbf{a}(\bq_{k+1}, t_{k+1}) \right) - \nonumber \\
		& \quad \frac{1}{2h} \nabla \mathbf{a}(\bq_k, t_k) \cdot (\bq_{k+1} - \bq_{k-1}) - \nabla \varphi(\bq_k, t_k),
\end{align}
for all $\bq_{k-1}, \bq_k, \bq_{k+1}$. However, investigation of the dependence on $\bq_{k+1}, \bq_{k-1}$ reveals that such a definition is possible only in the case of constant magnetic field. The left hand side of this equation is linear in the quantities $\bq_{k-1}, \bq_{k+1}$ and independent of $t_{k-1}, t_{k+1}$, so the right hand side must depend on these variables in the same manner to be satisfied at all $\bq_{k-1}, \bq_k, \bq_{k+1}, t_{k-1}, t_k, t_{k+1}$. This implies that the magnetic vector potential $\mathbf{a}(\bq,t)$ is linear in its spatial dependence and independent in time, and therefore, $\mathbf{B}(\bq,t)$ is constant. Of course, the case of constant magnetic field indicates that there are no higher-order terms in the Taylor expansion, and that no approximation is performed in Eq.\,(\ref{eq:vector_potential_approximation}).

\section{Conclusion}
\label{sec:conclusion}

In summary, the arguments delivered in Ref.\,\cite{Webb_2014} endeavor to establish the popular Boris scheme as a variational integrator, contradicting claims to the contrary in Refs.\,\cite{Qin_2013, Zhang_2014}. Because the arguments in Refs\,\cite{Qin_2013, Zhang_2014} do not rule out the existence of a discrete variational formulation for the Boris algorithm, we have furnished a proof that no such formulation exists based on a discrete Helmholtz condition. The establishment of the discrete variational formulation of the Boris scheme in Ref.\,\cite{Webb_2014} falls short due to the incomplete specification of the discrete Euler-Lagrange equations and (primarily) the introduction of truncation error after the action minimization procedure. The same critiques pertain to the parallel arguments for the relativistic charged particle algorithm in Ref.\,\cite{Webb_2014}. The good long-term behavior should then be attributed to the volume-preserving properties of the Boris scheme. 


\section{Acknowledgment} 
This work was supported by DOE contract number DE-AC02-09CH11466. We are especially grateful to the Referee who directed us to the discrete Helmholtz condition. 

\bibliographystyle{elsarticle-num}
\bibliography{SI_refs}

\begin{thebibliography}{10}
\expandafter\ifx\csname url\endcsname\relax
  \def\url#1{\texttt{#1}}\fi
\expandafter\ifx\csname urlprefix\endcsname\relax\def\urlprefix{URL }\fi
\expandafter\ifx\csname href\endcsname\relax
  \def\href#1#2{#2} \def\path#1{#1}\fi

\bibitem{Boris_1970}
J.~P. Boris, Relativistic plasma simulation - optimization of a hybrid code,
  in: Proceedings of the 4th Conference on the Numerical Simulation of Plasmas,
  1970, pp. 3--67.

\bibitem{Birdsall_1985_Boris}
C.~Birdsall, A.~Langdon, Plasma Physics Via Computer Simulation, McGrow-Hill
  Inc., 1985.

\bibitem{Stoltz_2002}
P.~H. Stoltz, J.~R. cary, G.~Penn, J.~Wurtele, Efficiency of a boris-like
  integration scheme with spatial stepping, Physical Review ST Accelerators and
  Beams 5 (2002) 094001.

\bibitem{Penn_2003}
G.~Penn, P.~H. Stoltz, J.~R. Cary, J.~Wurtele, Boris push with spatial
  stepping, Journal of Physics G: Nuclear and Particle Physics 29 (2003) 1719.

\bibitem{Qin_2013}
H.~Qin, S.~Zhang, J.~Liu, Y.~Sun, W.~M. Tang, Why is boris algorithm so good?,
  Physis of Plasmas 20 (2013) 084503.

\bibitem{Hairer_2006}
E.~Hairer, C.~Lubich, G.~Wanner, Geometric Numerical Integration, Springer,
  2006.

\bibitem{Xia_1992}
Z.~Xia, Existence of invariant tori in volume-preserving diffeomorphisms,
  Ergotic Theory and Dynamical Systems 12 (1992) 621--631.

\bibitem{Zaijiu_1994}
S.~Zaijiu, On the construction of the volume-preserving difference schemes for
  source-free systems via generating functions, Journal of Computational
  Mathematics 12 (1994) 265--272.

\bibitem{Quispel_1995}
G.~R.~W. Quispel, Volume-preserving integrators, Physics Letters A 206 (1995)
  26--20.

\bibitem{Quispel_1998}
G.~R.~W. Quispel, C.~P. Dyt, Volume-preserving integrators have linear error
  growth, Physics Letters A 242 (1998) 25--30.

\bibitem{He_2015}
Y.~He, Y.~Sun, J.~Liu, H.~Qin, Volume-preserving algorithms for charged
  particle dynamics, Journal of Computational Physics 281 (2015) 135--147.

\bibitem{Webb_2014}
S.~D. Webb, Symplectic integration of magnetic systems, Journal of
  Computational Physics 270 (2014) 570--576.

\bibitem{Marsden_2001}
J.~E. Marsden, M.~West, Discrete mechanics and variational integrators, Acta
  Numerica (2001) 1--158.

\bibitem{Zhang_2014}
S.~Zhang, Y.~Jia, Q.~Sun, {Comment on ``Symplectic integration of magnetic
  systems" by Stephen D. Webb [J. Comput. Phys. 270 (2014) 570-576]}, Journal
  of Computational Physics 282 (2014) 43--46.

\bibitem{Bourdin_2013}
L.~Bourdin, J.~Cresson, Helmholtz's inverse problem of the discrete calculus of
  variations, Journal of Difference Equations and Applications 19~(9) (2013)
  1417--1436.
\newblock \href {http://dx.doi.org/10.1080/10236198.2012.754435}
  {\path{doi:10.1080/10236198.2012.754435}}.

\bibitem{Helmholtz_1887}
H.~Helmholtz, {\"Uber die physikalische Bedeutung des Prinzips der kleinsten
  Wirkung}, J. Reine Angew. Math. 100 (1887) 137--166.

\bibitem{Douglas_1941}
J.~Douglas, Solution of the inverse problem of the calculus of variations,
  Trans. Amer. Math. Soc. 50 (1941) 137--166.

\end{thebibliography}

\end{document}